\documentclass{svjour2}
\usepackage{graphicx}

\def\beq{\begin{equation}}
\def\eeq{\end{equation}}
\def\rmd{{\rm d}}
\def\rmD{{\rm D}}

\journalname{General Relativity and Gravitation}

\begin{document}

\title{Fermi coordinates in Schwarzschild spacetime: closed form expressions}

\author{Donato Bini \and
         Andrea Geralico \and
        Robert T. Jantzen 
}

\institute{
              Donato Bini 
              \at
              Istituto per le Applicazioni del Calcolo ``M. Picone,'' CNR, I-00161 Rome, Italy\\
              ICRA, University of Rome ``La Sapienza,'' I--00185 Rome, Italy\\
              INFN - Sezione di Firenze, Polo Scientifico, Via Sansone 1, I--50019, Sesto Fiorentino (FI), Italy\\
              \email{binid@icra.it} 
\and
              Andrea Geralico 
              \at
              Physics Department and ICRA, University of Rome ``La Sapienza,'' I--00185 Rome, Italy\\
              \email{geralico@icra.it}   
\and
              Robert T. Jantzen
              \at
              Department of Mathematical Sciences, Villanova University, Villanova, PA 19085, USA
              and ICRA, University of Rome ``La Sapienza,'' I--00185 Rome, Italy\\
              \email{robert.jantzen@villanova.edu}
}

\date{Received: date / Accepted: date / Version: date}

\maketitle

\begin{abstract}
Fermi coordinates are constructed as exact functions of the Schwar\-zschild coordinates around the world line of a static observer in the equatorial plane of the Schwarzschild spacetime modulo a single impact parameter determined implicitly as a function of the latter coordinates. This illustrates the difficulty of constructing explicit exact Fermi coordinates even along simple world lines in highly symmetric spacetimes.

\keywords{Fermi coordinates \and Schwarzschild spacetime}
\PACS{04.20.Cv}
\end{abstract}

\section{Introduction}

An observer in arbitrary motion following a timelike world line in a given gravitational field measures that field and the nearby spacetime geometry by constructing a local coordinate system about that world line which acts as a ``laboratory reference system.''  
A Fermi coordinate system $(T,X^i), i=1,2,3$ introduced by Fermi \cite{fermi,bobfermi,manasse}
and named after him by Synge \cite{synge} is the mathematical realization of this idea.
Covering a small spacetime region around the world line at the origin $X^i=0$ of the spatial coordinates, 
the $X^i$ coordinate lines are spacelike geodesics orthogonal to the observer world line, and extend from an orthonormal triad of vectors in the local rest space of the observer which
form the observer's ``proper reference frame," discussed at length by Misner, Thorne and Wheeler \cite{MTW}, while $T$ represents the proper time along the observer world line. 

The geometry of such a coordinate system is very simple to describe, but very difficult to actually implement exactly and explicitly except for extremely simple world lines in very special spacetimes. For this reason most studies of Fermi coordinates in spacetime applications incorporate a series expansion approximation from the very beginning.
Fermi coordinates have been explicitly constructed in closed form around very special world lines in the de Sitter and G\"odel spacetimes by Chicone and Mashhoon \cite{chicmash}, using analytic expressions for the geodesics which exist in closed form and allow the Fermi coordinates to be expressed in terms of the original coordinate systems in which these spacetimes are usually described, as well as invert those relationships in closed form to express the latter coordinates in terms of the Fermi coordinates.
Chicone and Mashhoon also discuss
the difficulties one encounters in attempting to construct explicit Fermi coordinates in black hole spacetimes, where the case of an observer in geodesic radial motion was considered \cite{chicmash}.
Further examples of exact Fermi coordinate systems have been considered very recently by Klein and Collas \cite{klein} in the case of the Einstein static universe and the constant density interior Schwarzschild spacetime with cosmological constant and by Klein and Randles \cite{klein2} for a class of Robertson-Walker spacetimes.

Various authors have constructed Fermi coordinates in black hole spacetimes using a series expansion approximation from the very beginning \cite{linet,linet2,bgj}, but it is a rather cumbersome approach leading to a zoo of coefficients at each order arising from various derivatives of the metric, which gets increasingly complicated as the order is increased. In the present discussion, we go as far as possible
for static world lines representing observers at rest in the equatorial plane of the Schwarzschild spacetime based on the well known explicit exact solution of the geodesic equations expressed in terms of elliptic functions \cite{hagihara}. This leads to exact expressions for the Fermi coordinates as functions of the Schwarzschild coordinates modulo the numerical solution of a single nonlinear equation for an impact parameter describing the spacelike geodesics.  The spatial Fermi coordinates are then simply Riemann normal coordinates \cite{MTW} at a fixed point of space within the 3-dimensional geometry of the constant time hypersurfaces. In the flat spacetime limit of Schwarzschild for zero central mass, the process can be completed exactly to give orthonormal Minkowski coordinates based on the rest observer world line, as carried out in section 3. Appendix A deals with the special case in which the spacelike geodesics lie entirely in the equatorial plane. Appendix B considers the inverse coordinate transformation which must be evaluated by power series expansion.

It should be noted that the explicit closed form solution of the Schwarzs\-child geodesics has been used by Kraniotis and Whitehouse to calculate the perihelion precession and the orbital characteristics of Mercury, including its generalization to take into account the contribution from the cosmological constant as well \cite{kraniotis1} and then extended to the Kerr case \cite{kraniotis2}.
Hackmann and L\"ammerzahl \cite{hackmann} used these same explicit solutions to discuss the Pioneer anomaly, and later extended the explicit geodesic solutions to higher-dimensional spacetimes \cite{hackmannetal}.

\section{Preliminaries}

In a generic spacetime in a region endowed with existing coordinates $x^\alpha$ ($\alpha=0,1,2,3$) in terms of which the metric is known explicitly, Fermi coordinates are constructed as follows.
Let $x^\alpha(\tau)$ describe an observer's world line parametrized by the proper time $\tau$ and let its timelike unit tangent $U^\alpha=\rmd x^\alpha/\rmd \tau$  be the observer 4-velocity and  $a(U)=DU/d\tau$ the observer's 4-acceleration.
Choose a spatial (i.e., orthogonal to $U$) frame  $\{F_i\}$, $i=1,2,3$ along this world line which undergoes Fermi-Walker transport
\beq
\frac{\rmD F_i}{\rmd \tau}-[F_i \cdot a(U)] U =0\,.
\eeq

\begin{figure} 
\typeout{*** EPS figure }
\begin{center}
\includegraphics[scale=.4]{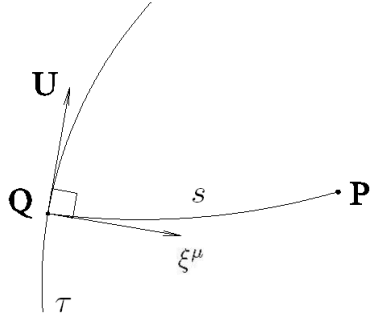}
\end{center}
\caption{
The construction of Fermi coordinates.} 
\label{fig:1}
\end{figure}

Consider all spacelike geodesics which pass through a generic point $Q$ on the observer's world line and which are orthogonal to $U$ at $Q$. These form a hypersurface, at least locally.
Let $P$  with coordinates $x^\alpha$ be a generic spacetime point near $Q$ on such a hypersurface and consider the unique spacelike geodesic segment from $Q$ to $P$ of proper length $s$.
The Fermi coordinates $(T,X,Y,Z)=(T,X^1,X^2,X^3)$ of $P$ are then defined by 
\beq
T=\tau\,, \qquad
X^i=s(\xi\cdot F_i)|_Q\,,
\eeq
where $\xi$ is the unit vector tangent to the spacelike geodesic segment at $Q$ satisfying the condition $(\xi \cdot U)|_Q=0$ (see Fig.~\ref{fig:1}).

The Schwarzschild metric expressed in standard coordinates $(t,r,\theta,\phi)$ is given by
\beq
\label{Schwmetric}
\rmd  s^2 = -N^2\rmd t^2 + N^{-2} \rmd r^2 
+ r^2 (\rmd \theta^2 +\sin^2 \theta\, \rmd \phi^2)\,,
\eeq
where the lapse function is 
$N=\left(1-{2M}/r\right)^{1/2}$ and it is understood that the following discussion only holds outside the horizon: $r>2M$ ($M$ is the mass parameter).
Introduce the usual orthonormal frame adapted to the static observers following the time lines
\beq 
\label{frame}
e_{\hat t}=N^{-1}\partial_t\,, \quad
e_{\hat r}=N\partial_r\,, \quad
e_{\hat \theta}={r}^{-1}\partial_\theta\,, \quad
e_{\hat \phi}=({r\sin \theta})^{-1}\partial_\phi\,.
\eeq

To construct Fermi coordinates along the world line of a static observer in this spherically symmetric spacetime,
it is enough to consider an observer at rest in the equatorial plane,  whose world line has the parametric equations $x^\alpha (\tau)$
\beq
\label{obswl}
t(\tau) = \frac{\tau}{N_0}\,, \quad 
r(\tau) = r_0\,, \quad 
\theta(\tau) = \frac{\pi}2\,, \quad 
\phi(\tau) = \phi_0\,, 
\eeq
where $\tau$ is the proper time parameter and $N_0=(1-2M/r_0)^{1/2}$ is the constant value of the lapse function along the world line. 
The 4-velocity of this observer is
$U={N_0}^{-1}\partial_t$
and its nonvanishing 4-acceleration is
\beq
a(U)=\nabla_U U=\frac{M}{N_0r_0^2}e_{\hat r}\,.
\eeq
Moreover, the spatial triad  
$\{e_{\hat r}$, $e_{\hat \theta}$, $e_{\hat \phi}\}$ 
consisting of the unit vectors associated with the spherical coordinate frame
form a Fermi-Walker dragged frame along the given world line of $U$.

Consider the geodesic equations in the Schwarzschild spacetime for affinely parametrized geodesics: $x=x(\lambda)$.
By virtue of the existence of the timelike and azimuthal Killing vectors one has constant energy ($E$) and angular momentum ($L$) per unit mass along each geodesic, leading to first integrals
\beq
\label{eqtphi}
\frac{\rmd t}{\rmd \lambda}= \frac{E}{N^2}, \qquad
\frac{\rmd \phi}{\rmd \lambda}=\frac{L}{r^2\sin^2\theta}\,.
\eeq
The remaining integrals of the geodesic equations governing radial and polar motion are given by 
\beq
\label{eqrth}
\frac{\rmd r}{\rmd \lambda} = \epsilon_r \sqrt{E^2-N^2\left(\mu^2+\frac{L^2+{\mathcal Q}}{r^2}\right)}\,, \quad
\frac{\rmd \theta}{\rmd \lambda} = \epsilon_\theta \frac{1}{r^2}\sqrt{{\mathcal Q}-L^2\cot^2\theta}\,,
\eeq
where ${\mathcal Q}\ge0$ (with ${\mathcal Q}=0$ only when $\theta=\pi/2$)
is a separation constant associated with a Killing tensor \cite{MTW} and $\mu^2=1,0,-1 $ for  timelike, null, spacelike  geodesics, respectively and $\epsilon_r=\pm 1$ and $\epsilon_\theta=\pm 1$ are sign indicators to keep track of all possible sign combinations.
By introducing the following convenient parametrization \cite{hughes}
\beq
\label{iJdefs}
L=J\cos i\,, \qquad {\mathcal Q}=J^2\sin^2i\,, \qquad L^2+{\mathcal Q}=J^2\,,
\eeq
with $J>0, i\in[0,\pi]$,
the equations governing spacelike geodesics ($\mu^2=-1$) become
\begin{eqnarray}
\label{eqsspat0}
\frac{\rmd t}{\rmd s} &=& \frac{E}{N^2}\,, \quad
\frac{\rmd r}{\rmd s} = \epsilon_r \sqrt{E^2+N^2\left(1-\frac{J^2}{r^2}\right)}\,, \nonumber\\
\frac{\rmd \theta}{\rmd s}&=&\epsilon_\theta \frac{J}{r^2}\sqrt{\sin^2i-\cos^2i\cot^2\theta}\,,\quad
\frac{\rmd \phi}{\rmd s} = \frac{J\cos i}{r^2\sin^2\theta}\,, 
\end{eqnarray}
where $s$ denotes the arclength parameter.

For the geodesics which are orthogonal to the static observer world line (\ref{obswl}), one must have $E=0$ simplifying the first two of the differential equations (\ref{eqsspat0}) to
\beq
\label{eqsspat}
\frac{\rmd t}{\rmd s} = 0\,, \quad
\frac{\rmd r}{\rmd s} = \epsilon_r \left( 1-\frac{2M}{r}\right)^{1/2} \left(1-\frac{J^2}{r^2}\right)^{1/2}\,,
\eeq
where $J$ is assumed to be larger than $2M$. 
Note that these geodesics are independent of the time coordinate and remain confined to the constant time hypersurfaces (which are extrinsically flat).
They must also satisfy $r\ge J$, and the value $r=J$ corresponds to a purely angular tangent vector, which occurs when $r$ equals the impact parameter $J$ for the geodesic, which is the minimum value of the radial variable along the geodesic.

For each value of $\tau$, let $x^i (s)$ be such an arclength parametrized spatial geodesic emanating from a point $Q$ on the static observer world line corresponding to the proper time $\tau$, i.e., $(x^i (0)) = (r_0,\pi/2,\phi_0)$. 
It has a unit tangent vector
\begin{eqnarray}
\xi(s) &=& \frac{\rmd x^\mu(s)}{\rmd s} \,\partial_\mu\nonumber\\
    &=& \epsilon_r \left(1-\frac{J^2}{r^2}\right)^{1/2} e_{\hat r}\nonumber \\
&&
  + \frac{J}{r}\left(\epsilon_\theta\sqrt{\sin^2i-\cos^2i\cot^2\theta} \, e_{\hat \theta}
  +\frac{\cos i}{\sin \theta}\, e_{\hat \phi}\right)
\,.
\end{eqnarray}
At $s=0$, this reduces to
\begin{eqnarray}
\label{xiQ}
\xi_Q \equiv
\xi(0) &=&
\epsilon_r \left(1-\frac{J^2}{r_0^2}\right)^{1/2} e_{\hat r}+\frac{J}{r_0}(\epsilon_\theta \sin i\, e_{\hat \theta}+\cos i\, e_{\hat \phi})\,, \nonumber\\
&=&
\epsilon_r \cos\chi_0 \,\, e_{\hat r}+\sin \chi_0\,(\epsilon_\theta \sin i\, e_{\hat \theta}+\cos i\, e_{\hat \phi})
\,,
\end{eqnarray}
where $r_0>J$ and the obvious notation
$\sin \chi_0={J}/{r_0}$ has been introduced. Thus apart from signs $(\chi_0,i)$ are spherical coordinates in the tangent space along the defining world line of the problem with the polar direction (from which $\chi_0$ is measured) aligned with the radial direction. The angle $i$ of the unit tangent about the initial radial direction in the equatorial plane also determines the plane of the geodesic, which is obtained by rotating the equatorial plane about that initial radial direction by this angle $i$, as a consequence of the spherical symmetry of the problem.
The parameter $J$ determines the initial angle $\chi_0$ of a geodesic required to hit a desired target point in the plane of the geodesic for a given value of the constant inclination angle $i$ of that plane about the radial direction. This aiming problem is the heart of the matter.

Fermi coordinates around the point $Q$ are then given by
\begin{eqnarray}
\label{fermicoords}
T&=&\tau = t\, (1-2M/r_0)^{1/2}
\,, \nonumber\\
X&=&s\, (\xi \cdot e_{\hat r})|_Q 
  =s \,\epsilon_r\, \left(1-\frac{J^2}{r_0^2}\right)^{1/2}\,,
\nonumber\\
Y&=&s\, (\xi \cdot e_{\hat \theta})|_Q
  = s \,\epsilon_\theta\, \frac{J\sin i}{r_0}\,,
\nonumber\\
Z&=&s\, (\xi \cdot e_{\hat \phi})|_Q
  = \,s\, \frac{J\cos i}{r_0} \,.
\end{eqnarray}
The arclength parameter $s$ as well as the orbital parameters $J$ and $i$ must be expressed in terms of the Schwarzschild coordinates $(r, \theta, \phi)$ with corresponding starting point $(r_0, \pi/2, \phi_0)$ by solving the geodesic equations. 

Because of the rotational symmetry about the initial radial direction, a geodesic starting with inclination angle $i$ about the radial direction will be confined to the plane through that radial direction with that inclination angle, so the geodesic problem is really a 2-dimensional one. One need only parametrize those geodesics by the polar angle $\alpha$ in that plane measured from the initial radial direction, and express the angular variables in terms of that angle, reducing the question of their motion to the relationship between $\alpha$ and $r$. This will be discussed in detail in the next two sections.

\section{The flat space case as a guide}

To clarify the more complicated situation of the Schwarzschild spacetime, we consider the flat spacetime limit $M=0$, where the Fermi coordinate system reduces to a new set of orthonormal Cartesian coordinates based at the point $Q$ with spherical coordinates $(r_0,\pi/2,\phi_0)$ and whose axes are aligned with the directions of the spherical orthonormal frame at that point. 
To avoid sign complications which lead to a nightmare of piecewise-defined functions,
we fix the base point $Q$ of the Fermi coordinate system to lie on the $x$-axis, namely with spherical coordinates $(r_0,\pi/2,0)$ and we assume that the point $P$ lies in the first octant with a larger radial coordinate, i.e., with spherical coordinates $(r,\theta,\phi)$ and  $r>r_0$, $0<\theta<\pi/2$, $0<\phi<\pi/2$. We also assume that passing from $Q$ to $P$ implies that the radial coordinate $r$ increases monotonically ($\epsilon_r=1$) and that the polar coordinate $\theta$ decreases monotonically ($\epsilon_\theta =-1$). As a consequence, 1) the point $H$ on the spacelike geodesic (straight line) connecting $Q$ to $P$ which lies at the minimal distance $J$ from the origin $O$ does not lie between $Q$ and $P$; 2) the inclination angle $i$ satisfies the relation $0<i<\pi/2$.
This situation for the point $H$ is illustrated in Fig.~2(a), in contrast with the case of Fig.~2(b) which will not be discussed here for simplicity. 

\begin{figure}[h] 
\typeout{*** EPS figure 2}
\begin{center}
\includegraphics[scale=0.3]{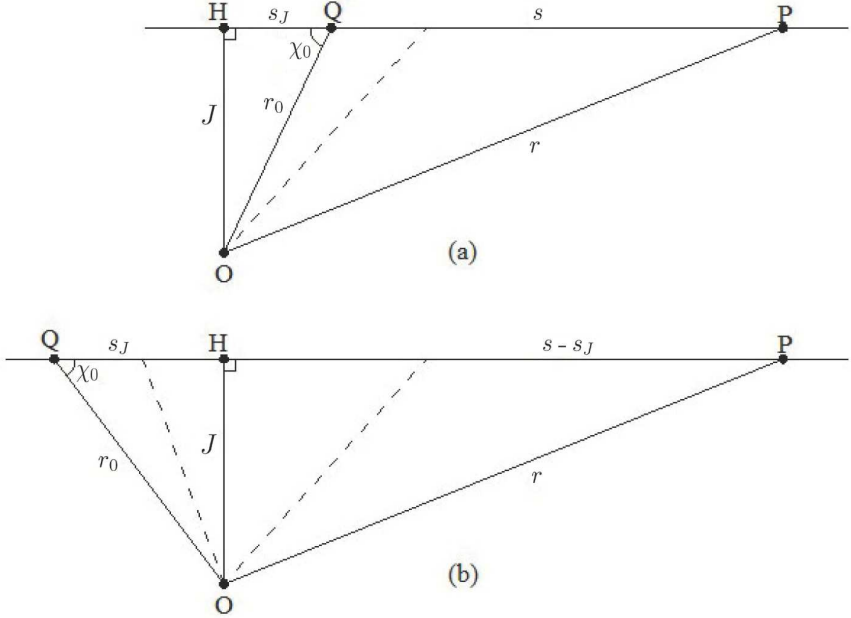}
\end{center}
\caption{
The geodesic straight line segment $QP$ in flat spacetime is shown in the two cases of $s_J<0$ (case (a): $Q$ between $H$ and $P$) and
$s_J>0$ (case (b): $Q$ not between $H$ and $P$). In the case  (a) we have $\epsilon_r=1$ since $r$ always increases starting from the initial value $r_0$;
In the case (b) we have instead  $\epsilon_r=-1$ as $r$ decreases  from the initial value $r_0$ down to the minimum value $J$, and then $\epsilon_r=1$ as $r$ increases  from  $J$ up to the generic value $r>J$.
} 
\label{fig:2}
\end{figure}

\begin{figure}[h] 
\typeout{*** EPS figure 3}
\vglue-0.15in
\begin{center}
\includegraphics[scale=0.3]{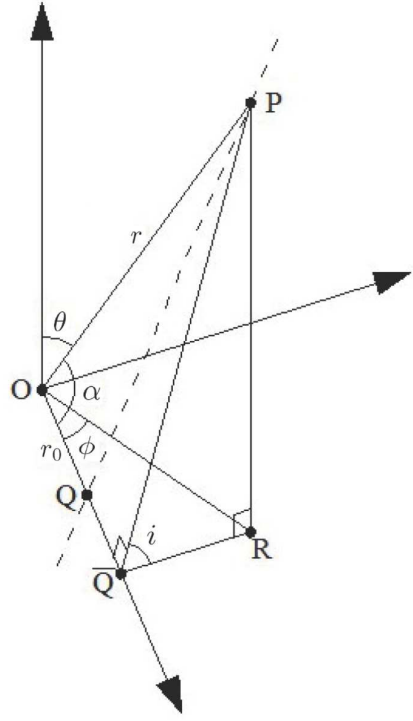}
\end{center}
\caption{The flat space geometry of a point $P$ along a geodesic segment $QP$ is shown in relation to the initial radial axis along $O\overline{Q}$ which contains the initial point $Q$ of the geodesic, with spherical coordinates ($r_0, \pi/2, 0$). The triangle $RO\overline{Q}$ lies in the equatorial plane $\theta=\pi/2$, the point $R$ is the projection of $P$ to the equatorial plane, and the point $\overline{Q}$ its projection along the initial radial axis. 
The  length of the line segment $PR$ is $r\cos \theta= r\sin\alpha \,\sin i$, leading to $\cos\theta=\sin\alpha\,\sin i$.
Similarly $\tan \phi=\displaystyle \frac{r\sin\alpha\,\cos i}{r\cos\alpha}=\tan\alpha\,\cos i$. Finally $r\cos\alpha=r\sin\theta\cos \phi$, which leads to  
$\cos\alpha=\sin\theta\cos \phi$.
} 
\label{fig:3}
\end{figure}

With this sign-fixing in mind, the explicit construction of a Fermi coordinate system around $Q$ proceeds straightforwardly.
The equations for spacelike geodesics orthogonal to the static observer world line (\ref{obswl}) reduce to
\begin{eqnarray}
\label{eqsspat1}
\frac{\rmd t}{\rmd s} &=& 0\,, \quad
\frac{\rmd r}{\rmd s} =  \sqrt{1-\frac{J^2}{r^2}}\,, \nonumber\\
\frac{\rmd \theta}{\rmd s}&=&  -\frac{J}{r^2}\sqrt{\sin^2i-\cos^2i\cot^2\theta}\,,\quad
\frac{\rmd \phi}{\rmd s} = \frac{J\cos i}{r^2\sin^2\theta}\,. 
\end{eqnarray}
The equation for $r$ can be easily integrated implicitly to yield
\beq
\label{s_sol_flat}
s= \sqrt{r^2-J^2}+s_J=\sqrt{r^2-J^2}-\sqrt{r_0^2-J^2}\,,
\eeq
or equivalently
\beq
\label{rquad}
r^2=r_0^2+s^2+2 s \sqrt{r_0^2-J^2}\,,
\eeq
which is equivalent to the law of cosines applied to the triangle $OQP$ in Fig. 2(a).
This relation can be solved for $J^2$ to yield
\beq
\label{J2_flat}
J^2=r_0^2-\frac{(r^2-r_0^2-s^2)^2}{4s^2}\,.
\eeq

The angular equations can be integrated by introducing the inclination angle $i$ of the plane of the triangle $POQ$ with respect to the horizontal and the new polar angular variable $\alpha$, as shown in Fig. 3. This leads to the equations
\beq\label{defalpha}
\cos \theta=\sin i\, \sin \alpha\,,\quad
\tan \phi = \cos i\,\tan \alpha \,.
\eeq
Our assumptions about the points $P$ and $Q$ imply $0< \alpha <\pi/2$.
Since the solution is independent of the symmetry angle $i$, both angular equations reduce to the following equation for the new angle $\alpha$
\beq
\frac{\rmd \alpha }{\rmd r}=\frac{J}{r\sqrt{r^2-J^2}}\,,
\eeq
whose solution is
\beq
\alpha= \arccos \left(\frac{J}{r}\right)-\arccos \left(\frac{J}{r_0}\right)\,,
\eeq
and therefore
\beq
\label{sinalpha_flat}
\sin \alpha =\frac{J}{r_0}\sqrt{1-\frac{J^2}{r^2}}-\frac{J}{r}\sqrt{1-\frac{J^2}{r_0^2}}
=\frac{J}{rr_0}\left(\sqrt{r^2-J^2}-\sqrt{r_0^2-J^2}\right)
\,.
\eeq
This can be re-expressed using Eq. (\ref{s_sol_flat}) to yield
\beq
\sin \alpha =\frac{ sJ}{rr_0}\,.
\eeq
Using Eq. (\ref{J2_flat}) to replace $J$, then $\cos \alpha$ can be re-expressed as
\begin{eqnarray}
\label{cosalpha_flat}
\cos \alpha&=&\sqrt{1-\frac{ s^2J^2}{r^2r_0^2}}=\frac{r^2+r_0^2-s^2}{2rr_0}\equiv \frac{\Omega}{2rr_0} \,,
\end{eqnarray}
where we have introduced the (positive) quantity
\beq
\Omega=r^2+r_0^2-s^2\,.
\eeq
From their definitions in Eq. (\ref{defalpha}) we then have 
\beq\label{costan}
\cos \theta =\sin i\,\displaystyle\frac{s J}{rr_0}\,,\quad
\tan \phi=  \cos i \frac{2 s J}{\Omega}
\,.
\eeq
Solving these for the angles and recalling the radial solution we have
\begin{eqnarray}
\label{finali}
r(s)&=& \sqrt{r_0^2+s^2+2s \sqrt{r_0^2-J^2}}\,,\nonumber \\
\theta (s)&=& \arccos \left(\frac{s J\sin i}{rr_0}\right)\,, \nonumber \\
\phi(s)&=& \arctan \left(\frac{2 s J\cos i }{r^2+r_0^2-s^2}\right)\,.
\end{eqnarray}

Equations (\ref{finali}) can be inverted to obtain $\{s,J,i\}$ as functions of $\{r,\theta,\phi\}$ and $\{r_0,\pi/2,0\}$. 
For this purpose it is convenient to introduce the quantity
\beq
\Sigma=\sin \theta \cos \phi\,.
\eeq
Now solving Eqs. (\ref{costan}) for $\sin i$ and $\cos i$
and eliminating $i$ using the fundamental trigonometric identity, one obtains
\beq
1=\sin^2 i+\cos^2 i
 =\frac{r^2r_0^2}{s^2J^2}\cos^2\theta +\frac{\Omega^2}{4s^2J^2}\tan^2 \phi \,.
\eeq
Next using the expression for $J$ given in Eq. (\ref{J2_flat}) leads to
\beq
\Omega =2rr_0  \Sigma   \,,
\eeq
or equivalently, from Eq. (\ref{cosalpha_flat})
\beq
\cos \alpha=\Sigma\,.
\eeq
From its definition $\Omega=r^2+r_0^2-s^2$ we then find
\beq
s=\sqrt{r^2+r_0^2-2  rr_0\Sigma } \,.
\eeq
Back-substituting the above expression for $s$ into Eq. (\ref{J2_flat}) determines $J$ and $sJ$ to be
\beq
J=\frac{rr_0\sqrt{1-\Sigma^2}}{\sqrt{r^2+r_0^2-2  rr_0 \Sigma}}\,,\qquad sJ=rr_0\sqrt{1-\Sigma^2}\,.
\eeq
Finally  inserting these values into Eq. (\ref{costan}) and solving for $\sin i$ and $\cos i$  one finds
\beq
\sin  i = \frac{\cos \theta}{\sqrt{1-\Sigma^2}}\,,\quad
\cos  i = \frac{\sin \theta \sin  \phi }{\sqrt{1-\Sigma^2}}\,.
\eeq
Summarizing we have
\begin{eqnarray}
s&=& \sqrt{r^2+r_0^2-2 rr_0 \Sigma }\,,\nonumber \\
J&=&\frac{rr_0\sqrt{1-\Sigma^2}}{\sqrt{r^2+r_0^2-2  rr_0  \Sigma }}\,,\nonumber \\
i &=& \arcsin \left( \frac{ \cos \theta  }{\sqrt{1-\Sigma^2}}\right)\,.
\end{eqnarray}

Note also that
\beq
1-\frac{J^2}{r_0^2}=\frac{( r \Sigma -r_0)^2}{s^2}\,.
\eeq
and hence
\beq
s\sqrt{1-\frac{J^2}{r_0^2}}=  r\Sigma -r_0 \,.
\eeq
The map between Fermi and spherical coordinates follows  then from Eqs. (\ref{fermicoords}) 
\begin{eqnarray}
\label{fermi2}
X=r\Sigma -r_0\,,\quad
Y=-r  \cos \theta \,,\quad
Z=r\sin \theta \,\sin \phi\,.
\end{eqnarray}

\section{The Schwarzschild case}

In the Schwarzschild case the radial behavior versus arclength along spatial geodesics also decouples from the angular motion.
To avoid sign complications, we continue to fix the base point $Q$ of the Fermi coordinate system on the $x$ axis, namely with spherical coordinates $(r_0,\pi/2,0)$ and we assume that the point $P$ lies in the first octant, i.e., with spherical coordinates $(r,\theta,\phi)$ and  $r>r_0$, $0<\theta<\pi/2$, $0<\phi<\pi/2$.
In addition we require now that $r_0>2M$ to avoid the coordinate singularity at $r_0 = 2M$. 
As before, we also assume that passing from $Q$ to $P$ implies that the radial coordinate $r$ increases monotonically ($\epsilon_r=1$) and that the polar coordinate $\theta$ decreases monotonically ($\epsilon_\theta =-1$), exactly as illustrated in Fig. 2(a) for the flat spacetime case, so that $0<i<\pi/2$.

We start now by integrating the radial equation to find the relationship of $s$ versus $r$
\beq
\frac{ds}{dr}=\frac{r^{3/2}}{\sqrt{(r^2-J^2)(r-2M)}} \,.
\eeq
The solution for $J\neq0$, namely
\begin{eqnarray}\label{sofr}
s=\int_J^r \frac{r^{3/2}}{\sqrt{(r^2-J^2)(r-2M)}} \, dr+s_J\,,
\end{eqnarray}
can be written as (see \cite{elliptic}, p. 130, Eq. 258.11 with $m=2$)
\begin{eqnarray}\label{sofr2}
s&=& A_E [E(\beta,k)-E(\beta_0,k)]+A_F[F(\beta,k)-F(\beta_0,k)]\nonumber \\
&&+A_\Pi[\Pi(\beta,n,k)-\Pi(\beta_0,n,k)]\nonumber\\
&&+\frac1N\sqrt{r^2-J^2}-\frac1{N_0}\sqrt{r_0^2-J^2}\,,
\end{eqnarray}
where $F$, $E$ and $\Pi$ are elliptic integrals of the first, second and third kind, respectively,
\beq
[A_E,A_F,A_\Pi]=\sqrt{J(J+2M)}\left[-1,\frac{J^2+4M^2}{J(J+2M)},\frac{2M}{J}\right]
\eeq
and
\begin{eqnarray}
\label{vincsubs}
\beta &=&\arcsin\sqrt{\frac{(J+2M)(r-J)}{2J(r-2M)}}\,,\qquad 
\beta_0 =\beta(r_0)\,,\nonumber\\
n&=&\frac{2J}{J+2M}\,,\qquad
k=2\sqrt{\frac{M}{J+2M}}\,.
\end{eqnarray}
Note that in the flat spacetime limit $M\to0$ we have
\beq
\label{vincsubs0}
\beta \to\arcsin\sqrt{\frac{r-J}{2r}}\,,\qquad 
n\to2\,,\qquad
k\to0\,,
\eeq
and
\beq
[A_E,A_F,A_\Pi]\to J\left[-1,1,0\right]\,,
\eeq
implying that the arclength function (\ref{sofr2}) reduces to Eq. (\ref{s_sol_flat}), since 
\beq
[E(\beta,k),F(\beta,k)]\to\arcsin\sqrt{\frac{r-J}{2r}}\,,\qquad
\Pi(\beta,n,k)\to{\rm arctanh}\sqrt{\frac{r-J}{r+J}}\,,
\eeq
and $[N,N_0]\to1$ (see \cite{elliptic}, p. 10, Eq. 111.01).

Moreover, to first order in $M$ we have
\beq
\frac{ds}{dr}= \frac{r}{\sqrt{r^2-J^2}}\left(1+\frac{M}{r}\right)+O(M^2)\,.
\eeq
with solution
\beq
 s(r)= s_J+\sqrt{r^2-J^2}+  M\ln(r+\sqrt{r^2-J^2})+O(M^2)\,,
\eeq
which shows the lowest order correction to the previous flat spacetime case.

The equations for the angular motion are formally the same as in the flat spacetime case
\beq
\label{eq_th_and_phi}
\frac{\rmd \theta}{\rmd s}= -\frac{J}{r^2}\sqrt{\sin^2i-\cos^2i\cot^2\theta}\,,\quad
\frac{\rmd \phi}{\rmd s} = \frac{J\cos i}{r^2\sin^2\theta}\,, 
\eeq
hence we may use the same  parametrization of the angular variables along the flat spacetime geodesics by the polar angle $\alpha$
\beq\label{zdef}
\cos\theta=\sin i \, \sin \alpha\,, \qquad \tan\phi = \cos i \, \tan\alpha
\eeq
with the same geometric interpretation.
Isolating $\sin i$ and $\cos i$ from Eqs. (\ref{zdef}) and using the trigonometric identity $\sin^2 i+\cos^2 i=1$  again yields 
\beq
\cos\alpha = \sin\theta \, \cos \phi =\Sigma
\eeq
and hence we find the same result as in the flat spacetime case 
\begin{eqnarray}
\label{isol}
\sin i &=&  \frac{\cos\theta}{\sqrt{1-\Sigma^2 }}\,,\quad
\cos i =\frac{ \sin \phi  \sin\theta}{\sqrt{1-\Sigma^2 }}\,. 
\end{eqnarray}

The angular equations (\ref{eq_th_and_phi}) reduce then to the single equation
\beq\label{eq:alpha_and_phi}
\frac{\rmd\alpha}{\rmd s} =  \frac{J}{r^2}\,, 
\eeq
from which it follows that
\beq
\label{eqalphadir}
\frac{d\alpha}{dr}
=\frac{J }{\sqrt{r(r-2M)(r^2-J^2)}}\,.
\eeq
The solution with $\alpha(r_0)=0$ (see \cite{elliptic}, p. 128, Eq. 258.00) is
\beq
\label{alpha_sol}
\alpha
=2\sqrt{\frac{J}{J+2M}}[F(\beta,k)-F(\beta_0,k)]\,,
\eeq
where $\beta$ and $k$ are given by Eq. (\ref{vincsubs}).
In the flat spacetime limit $M\to0$ we recover Eq. (\ref{sinalpha_flat}). In fact, the previous equation becomes
\beq
\alpha
=2\left(\arcsin\sqrt{\frac{r-J}{2r}}-\arcsin\sqrt{\frac{r_0-J}{2r_0}}\right)\,,
\eeq
whence
\beq
\cos\frac{\alpha}2=\frac1{2\sqrt{rr_0}}(\sqrt{(r+J)(r_0+J)}+r-J)\,,
\eeq
which immediately gives Eq. (\ref{sinalpha_flat}) by simple trigonometric relations. 

It then follows that
\beq
 \sin\theta \, \cos \phi 
 = \cos \left(2\sqrt{\frac{J}{J+2M}}[F(\beta,k)-F(\beta_0,k)]\right)\,.
\eeq
This equation {\it implicitly} determines the impact parameter $J$ as a function of the Schwarzschild coordinates of the points $Q$ and $P$. In other words, 
once the coordinates of the points $Q$ and $P$ are fixed, then $J$ is found numerically and, consequently, the arclength function $s$ is determined by Eq.~(\ref{sofr}).
This completely expresses the Fermi coordinates (\ref{fermicoords}) exactly but implicitly as functions of the Schwarzschild coordinates at least locally where this admits a solution. 

This is the key point of the present work and pinpoints the difficulty in general of finding exact expressions for the Fermi coordinates in terms of the original Schwarzschild coordinates.
The inverse map from Fermi to Schwarzschild coordinates cannot be obtained explicitly, but only as a series expansion as  briefly discussed in Appendix B.

\section{Concluding remarks}

Fermi coordinates are widely used in the literature since they provide a continuous locally inertial coordinate system along the world line of an observer allowing physical measurements in a gravitational field to be expressed simply in those coordinates. However, it is difficult in practice to actually represent Fermi coordinates in terms of symmetry adapted coordinates even along simple world lines in highly symmetric spacetimes. In the Schwarzschild case along static world lines this is facilitated by the decoupling of the radial and polar angular geodesic motion, which is broken as soon as one considers repeating this analysis for a Kerr black hole. One succeeds only in expressing the Fermi coordinates as functions of the Schwarzschild coordinates using an implicitly determined impact parameter for the spatial geodesics.
In any case the reverse coordinate transformation resists analytic treatment but at least in the Schwarzschild case under consideration, expressing it using a series expansion can be slightly facilitated by the exact geodesic solutions in parametric form.

\appendix

\section{Equatorial plane geodesics}
\label{special}

The equatorial plane case $Q=0$ presents two exceptions to the general discussion, i.e., the curves emanating from the world line of the construction in the equatorial plane can be either radial  or nonradial equatorial plane geodesics. Even if in this case the discussion can be done in general, for simplicity we continue to restrict the discussion to the first octant, which is now the first quadrant of the equatorial plane.

\subsection{Radial equatorial plane geodesics}

The outgoing radial geodesics ($\phi=0$, $\theta=\pi/2$, $\epsilon_r=1$) correspond to the case $J=0=L$
and can be integrated trivially to obtain 
\beq
s(r)=(rN-r_0N_0)+M\ln\left(\frac{r-M+rN}{r_0-M+r_0N_0}\right)\,. 
\eeq
The transformation from Schwarzschild to Fermi coordinates is given by
\beq
T= t\, N_0\,, \quad
X=s(r)\,, \quad
Y=0=Z\,.
\eeq

\subsection{Nonradial equatorial plane geodesics}

The nonradial equatorial plane geodesics ($\theta=\pi/2$) correspond to the case $J=L\neq0$ but $i=0$.
The geodesic equations reduce to 
\beq
\label{equateqs}
\frac{\rmd t}{\rmd s} = 0\,, \qquad
\frac{\rmd r}{\rmd s} =  \left(1-\frac{2M}{r}\right)^{1/2} \left(1-\frac{J^2}{r^2}\right)^{1/2}\,, \qquad
\frac{\rmd \phi}{\rmd s} = \frac{J}{r^2}\,.
\eeq
Fermi coordinates are 
\beq
T= t\, (1-2M/r_0)^{1/2}\,, \quad
X=s (1-J^2/r_0^2)^{1/2}\,, \quad
Y=0\,, \quad
Z=s J/r_0\,,
\eeq
where $s$ is still given by Eq.~(\ref{sofr}).

The orbits can be parametrized by the azimuthal angle $\phi$ according to
\beq
\label{r_di_phi_equat}
\frac{\rmd r}{\rmd \phi} = \frac{r^2}{J}\left(1-\frac{2M}{r}\right)^{1/2}\,\left(1-\frac{J^2}{r^2}\right)^{1/2}\,,
\eeq
whose solution is
\beq
\label{solrequat}
r =\frac{2M}{4\wp(\phi+c,g_2,g_3)+1/3}\,,
\eeq
where $\wp$ is the Weierstrass elliptic function \cite{abrasteg} and
\beq
\label{g23sol}
g_2=\frac{1}{\bar J^2}+\frac{1}{12}\,, \quad
g_3=\frac{1}{216}-\frac{1}{6\bar J^2}\,,\quad 9(g_2+6g_3)=1\,,
\eeq
with $\bar J=J/M$.
The integration constant $c$ is chosen such that  
\beq
r_0=\frac{2M}{4\wp(c,g_2,g_3)+1/3}\,.
\eeq
Finally Eq.~(\ref{solrequat}) determines $J$ implicitly as a function $r$, $r_0$, $\phi$ and $\phi_0$.
Therefore, no significant simplifications arise in this case with respect to the general situation.
Note that replacing $\phi$ by $\alpha$ leads to the general relationship between $r$ and $\alpha$ off the equatorial plane.
Inverting this relation (\ref{solrequat}) with this substitution leads  to Eq. (\ref{alpha_sol}) in the main text.

\section{Inverse transformation: from Fermi to Schwarzschild}
\label{seriesalpha}

The inverse map from Fermi to Schwarzschild coordinates can be obtained only as a series expansion.
A convenient procedure for evaluating this expansion is outlined below.

Consider Eq. (\ref{eqalphadir}) 
\beq
\label{intermsofz}
\frac{\rmd r}{\rmd \alpha} =  \frac{r^2}{J}\left(1-\frac{2M}{r}\right)^{1/2}\, \left(1-\frac{J^2}{r^2}\right)^{1/2}\,,
\eeq
whose solution is 
\beq
r(\alpha)=\frac{2M}{4\wp(\alpha+c,g_2,g_3)+1/3}\,,
\eeq
where $g_2$ and $g_3$ are given by Eq.~(\ref{g23sol}),
the constant $c$ being determined by requiring that $r=r_0$ when $\alpha=0$, i.e.
\beq
c=\wp^{-1}\left(\frac{r_0-6M}{12r_0},g_2,g_3\right)\,.
\eeq

Consider then a series solution for Eq.~(\ref{eq:alpha_and_phi}), i.e., 
\beq
\alpha=\frac{J}{r_0^2}\,s+\frac{J^2}{r_0^4}N_0\cot\chi_0\,s^2+O(s^3)\ ,
\eeq
up to second order in $s$.

The inverse transformation expressing the Schwarzschild coordinates in terms of the Fermi coordinates is thus given by 
\begin{eqnarray}
\label{fermitoschw}
r&=&\frac{2M}{4\wp(\alpha+c)+1/3}\,,\nonumber \\
\theta&=&\arccos(\sin i\sin \alpha)\,,\nonumber \\
\phi&=& \arctan(\cos i\tan \alpha)\,.
\end{eqnarray}
The final step consists of eliminating the parameters of the orbit by using Eq.~(\ref{fermicoords}), i.e.
\beq 
\cos\chi_0\to\frac{X}{s}\,, \quad 
\sin i\to\frac{Y}{s\sin\chi_0}\,, \quad 
\cos i\to\frac{Z}{s\sin\chi_0}\,, \quad
\sin\chi_0\to\frac{\sqrt{Y^2+Z^2}}{s}\ ,
\eeq 
also taking into account the relations $J=r_0\sin \chi_0$ and $s=\sqrt{X^2+Y^2+Z^2}$.
One then obtains
\beq
\sin i=\frac{Y}{\sqrt{Y^2+Z^2}}\ , \quad
\cos i=\frac{Z}{\sqrt{Y^2+Z^2}}\ , \quad
J=r_0\frac{\sqrt{Y^2+Z^2}}{\sqrt{X^2+Y^2+Z^2}}\ ,
\eeq
allowing us to express all the Schwarzschild coordinates (\ref{fermitoschw}) only in terms of their initial values and Fermi coordinates.
For instance, up to second order in the spatial Fermi coordinates we have
\begin{eqnarray}
\label{schwtofermi}
r&=&r_0+N_0 X+\frac12\left[\frac{M}{r_0^2}X^2+\frac{N_0^2}{r_0}(Y^2+Z^2)\right]+O({\rm Fermi}^3)\,, \nonumber\\
\theta&=&\frac{\pi}{2}+\frac{Y}{r_0}-\frac{N_0}{r_0^2} XY+O({\rm Fermi}^3)\,, \nonumber\\
\phi&=&\phi_0+\frac{Z}{r_0}-\frac{N_0}{r_0^2} XZ+O({\rm Fermi}^3)\,,
\end{eqnarray}
a result originally obtained by Leaute and Linet \cite{linet} and later generalized to the case of a static observer located at any point on the equatorial plane of the Kerr spacetime and to any uniformly rotating circular equatorial orbit by Bini, Geralico and Jantzen \cite{bgj}.
Higher order terms can be obtained straightforwardly in this way.

\end{document}